\documentclass[a4paper]{jpconf}
\usepackage{epsfig}
\usepackage{dcolumn}
\usepackage{epstopdf}
\usepackage[T1]{fontenc}
\usepackage[latin9]{inputenc}
\usepackage{amstext}
\usepackage{graphicx}

\providecommand{\tabularnewline}{\\}

\begin{document}

\title{Multiferroicity in V-doped PbTiO$_{3}$}

\author{Francesco Ricci, Vincenzo Fiorentini}

\address{CNR-IOM and Dept. of Physics, University of Cagliari, Cittadella Universitaria, 09042 Monserrato (CA), Italy}
\ead{frankyricci@gmail.com, vincenzo.fiorentini@gmail.com}

\begin{abstract}
We report \emph{ab initio} predictions on the proper multiferroic (ferromagnetic, insulating and ferroelectric) character of PbTiO$_{3}$ doped with vanadium. 
V impurities coupled ferromagnetically  carry a magnetization of 1 $\mu_{\rm B}$ each. The coupling is expected to be strong, since the paramagnetic solution is higher by 150 meV/vanadium, and no stable antiferromagnetic solution was found. The electronic gap in the doped system is about 0.2-0.3 eV in GGA, hence the system is properly multiferroic. V doping increases the spontaneous polarization in PbTiO$_{3}$,  with an approximate percentual rate of  0.7 $\mu$C/cm$^{2}$.\\

\end{abstract}

\section{Introduction}

Modern technology uses extensively ferroic materials --i.e. having a spontaneous and permanent order parameter of one kind or another-- with fairly exotic properties. For example, commodity data-storage devices exploit ferromagnets with giant magnetoresistance in magnetic random-access
memories, in a continued push toward increased information density and reduced dimensions and writing energy.
 Besides their traditional use in sensors and actuators, ferroelectrics --with their permanent and switchable electrical
polarization-- are the building block of non-volatile, high speed, random-access
memories  which promise  performances superior  to semiconductor flash memories. 

While their  existence has been known since the 1960s, multiferroics, i.e. materials where polarization P and magnetization M coexist \cite{review1,review2},  have enjoyed a renaissance in recent years, because of great
improvements in growth techniques and new theoretical approaches. 
The coexistence of several order parameters and their mutual coupling may  open the way to new device concepts,   including the  electrical addressing of  magnetic memories without magnetic fields and their  generating  currents; the
creation  of multi-layer multi-state logical devices, exploiting combinations of 
polarization and magnetization; and magnetoelectric sensors.
However, most of the current multiferroics lack sufficiently strong polarization or magnetization, or sufficient magnetoelectric coupling, or the correct kind of magnetic order (ferromagnets are desired, but antiferromagnets are much more common). 

Some of the difficulties of single-phase multiferroics  are bypassed by ``metamultiferroics'' built of multilayer heterostructures of ferromagnets and ferroelectrics \cite{loro}, as for example  in   multiferroic tunnel junctions (MFTJ), where ferromagnetic electrodes are separated by a ferroelectric insulating barrier. The  resistance against tunneling across the junction depends on the relative orientation of magnetization of the two electrodes (magnetoresistance) and on the direction of ferroelectric polarization of the insulating layer (electroresistance). This enables  in principle a 4-level switchable resistance using external electric and magnetic fields. 
Of course, a further magnetic degree of freedom in the insulating-barrier material could potentially duplicate the number of states, and any chance of locally tuning the magnetization would add further design leeway. Following this line of thought, in this paper we report an \emph{Ab initio} investigation strongly suggesting that  PbTiO$_{3}$ (PTO) doped with V (V:PTO) is a ferromagnetic small-gap insulator, hence a proper multiferroic. This is particularly interesting in view of the giant electroresistance  effect predicted in tunnel junctions with PTO as ferroelectric tunnel layer \cite{noi}.

\section{V doping of PTO}

Here  we report the first principles prediction of a multiferroic
state of lead titanate  doped with magnetic vanadium. PTO is a well-known tetragonal perovskite with a high spontaneous
polarization (86 $\mu$C/cm$^{2}$) of displacive origin. PbVO$_{3}$ also happens to be tetragonal, and with an even higher spontaneous polarization (152 $\mu$C/cm$^{2}$).
Previous  studies have shown its ground state to be an antiferromagnetic
insulator (C-type), making it not especially interesting as a multiferroic.
Motivated by our previous study of V-doped ferroelectric titanates \cite{noi2}, 
we examine  the magnetic properties of vanadium diluted within the robust PTO ferroelectric.

\emph{Ab initio} calculations  are performed within 
density functional theory in the generalized
gradient approximation (GGA) by Perdew-Wang  
using the PAW method \cite{paw} 
as implemented in the VASP code \cite{vasp}. Standard cutoff is used for the plane wave basis, and the k-point mesh for the bulk is 8$\times$8$\times$8 (appropriately rescaled  for defect super cells). The Berry phase technique is used to calculate the polarization change upon V doping, using strings of 16 points in the polarization direction.

\begin{figure}[ht]
\begin{centering}
\centerline{\includegraphics[scale=0.9]{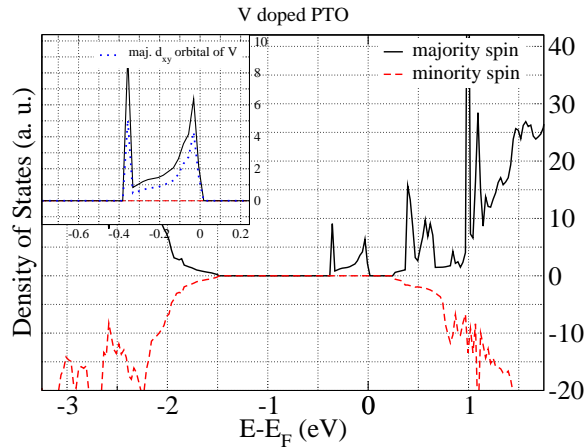}}
\end{centering}
\caption{\label{fig:dos}Density of states for V:PTO. Positive curve
is  majority spin, negative is minority spin. Fermi energy is at 0 eV. Inset shows a zoom of the gap peak. The dotted line is a  V $d$-up.}
\end{figure}

After studying PTO and PVO in the respective ferroelectric phases, we substituted V for one Ti in a 2$\times$2$\times$2 tetragonal supercell of ferroelectric PTO, i.e. dopant concentration $\sim$12.5\%. A selection of structure parameters (lattice constants, distance of V or Ti from neighbouring O along the polar axis) are collected in Tab. \ref{tab:table}. V has one excess electron compared to Ti, so it is, as expected, spin-polarized with a moment of 1 $\mu_{\rm B}$. Orbital angular  momentum is assumed to be quenched.
Fig. \ref{fig:dos} displays the key result, i.e. the density of states (DOS) of ferromagnetic V:PTO. As we can observe, a fully occupied peak appears in the gap of PTO, placed so that  a gap of about 0.2 eV survives. This DOS feature is obviously related to the excess V electron occupying antibonding conduction states. Indeed, as shown in the inset, the main contribution to that peak is 
that of spin-up \emph{d} orbitals of V. Thus,  PTO doped with vanadium is a small-gap  ferromagnetic insulator thanks to the unpaired electron of dopant vanadium. The gap of PTO is of course underestimated due to the known \cite{gap-err} gap error of semi-local functionals; it may well be that the small V-related gap be also somewhat larger than calculated due to the same effects  (self-interaction, xc discontinuity, etc.).

\begin{table}
\begin{centering}
\begin{tabular}{cccc}
\hline 
 & V:PTO & PTO (Theor. / Expt.) & PVO (Theor. / Expt.)\tabularnewline
\hline 
a & 7.848 & 3.924 / 3.895  & 3.806 / 3.804\tabularnewline
\hline 
c & 8.351 & 4.175 / 4.171 & 4.979 / 4.677\tabularnewline
\hline 
V-O$_{\rm top}$ & 1.709 & - & 1.700 / 1.677\tabularnewline
\hline 
V-O$_{\rm down}$ & 2.377 & - & 3.278 / 3.01\tabularnewline
\hline 
Ti-O$_{\rm top}$ & 1.792 & 1.795 / 1.75 & -\tabularnewline
\hline 
Ti-O$_{\rm down}$ & 2.380 & 2.380 / 2.42 & -\tabularnewline
\hline 
\end{tabular}
\par\end{centering}
\caption{\label{tab:table}Cell parameters for V doped PTO, bulk PTO and PVO. 
Theoretical values are calculated for this work in GGA approximation.
Experimental values are from Ref. \cite{key-1,key-2}}
\end{table}

In the cell just discussed, V is ferromagnetic by construction, being coupled to its periodic images. To study an antiferromagnetic configuration we consider two V's at the same concentration in a  2$\sqrt{2}$$\times$2$\sqrt{2}$$\times$2 tetragonal cell. The two V are placed in that cell as far as possible from each other. While the  
ferromagnetic configuration is stable, the antiferromagnetic
configuration is not. The V moments disappear and the system converges to a metallic Pauli paramagnet of zero moment, which is about 300 meV higher in energy than the ferromagnet. Of course, this prevents the evaluation of a coupling parameter for magnetic models, but the ferromagnet seems quite stable nevertheless.

\begin{figure}[ht]
\centerline{\includegraphics[scale=0.4]{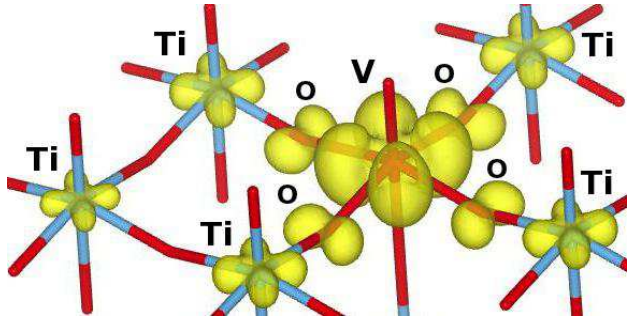}
}
\caption{Majority spin density of the spin-polarized electron of V in the basal plane (iso-level: 10$^{-3}$\,$e$/\AA$^{3}$). The state is anti-bonding with first-neighbor O's, has $d_{xy}$-like nodal structure, and propagates through $d_{xy}$-like Ti states. Charge paths along the vertical axis are  negligible.}
\end{figure}

Inspection of the charge density of the impurity state (Fig. 2) suggests a $d_{xy}$-like nodal structure and coupling of V's via V-O-Ti-O-... paths in the basal plane, whereas hardly any density is to be found along the vertical axis. While  the V-centered state is, as expected, anti-bonding with first neighbors, the bonding O-V  states (which mimic those of the substituted Ti) appear to make for a larger charge accumulation in the V-O bond region (Fig. 3), matching the shorter V-O bond length compared to Ti-O (1.71 vs 1.79 \AA).

\begin{figure}[ht]
\centerline{\includegraphics[scale=0.4]{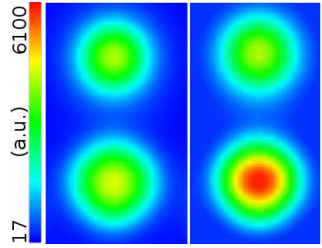}}
\caption{Total charge density (in arbitrary units) in the vertical plane across a Ti-O bond in PTO (left) and a V-O bond in V:PTO. A slightly larger bond charge can be appreciated in V:PTO.}
\end{figure}

Following Ref.\cite{key-3}, the equilibrium concentration is $\left[\rm V\right]$=$N_{s}\exp\left(-\rm E_{f}\left(\rm V\right)/kT_{g}\right)$,
where $N_{s}$ is the  density of Ti sites and $T_{g}$ the growth temperature. Assuming that the solubility limit for V in PTO is PVO, i.e. that high V contents will eventually cause phase separation of PVO, the formation energy is
E$_{f}$(V)=E(V:PTO)--($n$--1)\,E(PTO)--E(PVO)$\approx$1.23 eV, involving the total energy of  the doped supercell
of V:PTO comprising $n$ unit cells, and of the unit cells of PTO and PVO.  Since $N_{s}$=1.25$\cdot$10$^{23}$cm$^{-3}$, at a typical $T_{g}$=650 K
we find a logarithmic concentration $\log_{10}\left[\rm V\right]\approx$13.4, which may increase perhaps to about 16 due to vibrational entropy (V is heavier than Ti). This 
thermodynamic solubility limit is relatively low, but it may well not apply whenever  kinetics or  non-equilibrium phenomena become dominant, as it happens in modern epitaxial growth techniques.

Finally, we estimated the changes in polarization induced by V doping. At 12.5\% V we find a polarization increase of 8.3 $\mu$C/cm$^{2}$, which agrees nicely with  8.25 $\mu$C/cm$^{2}$  obtained by linear interpolation between the bulk values of PTO and PVO  mentioned previously. Therefore,  in terms of polarization, V does not seem detrimental, but in fact benign.

\section{Conclusions}

In summary, we presented an \emph{Ab initio} prediction of the proper  multiferroicity of V-doped PTO.  The gap is about 0.2-0.3 eV and the magnetization is 1 $\mu_{\rm B}$/V, or 1.5$\times$10$^{22}$ spins/cm$^{3}$ at our chosen V density. We predict the polarization to increase linearly with V concentration at the percentual rate of 0.7 $\mu$C/cm$^{2}$. Presumably this will only apply at relatively low concentrations, since the  antiferromagnetic character of PVO will eventually take over.\\

\noindent {\it Note} -- During review we have become aware of a very recent  study \cite{indiani} of transition-metal doping in a perovskite ferroelectric (BaTiO$_{3}$), reporting results in general agreement with the ones just presented.


\section*{References}
\begin{thebibliography}{References}

\bibitem{review1}
K.F. Wang, J.-M. Liu, and Z.F. Ren, Advances in Physics Vol. {\bf 58}, No. 4, 321-448 (2009).

\bibitem{review2}
J. P. Velev, S. S. Jaswal, and E. Y. Tsymbal, Phil. Trans. R. Soc. A {\bf 369}, 3069-3097 (2011).


\bibitem{loro}
J. P. Velev, C.-G. Duan, J. D. Burton, A. Smogunov, M. K. Niranjan, E. Tosatti, S. S. Jaswal, and E. Y. Tsymbal, Nano Lett. {\bf 9}, 427 (2009)


\bibitem{noi} 
F. Ricci, A. Filippetti, and V. Fiorentini, to be published.

\bibitem{noi2}
M. Scarrozza, A. Filippetti, and V. Fiorentini,
Phys. Rev. Lett. {\bf 109}, 217202 (2012).

\bibitem{paw}
P. E. Bl\"ochl, Phys. Rev. B {\bf 50}, 17953 (1994); G. Kresse and D. Joubert, Phys. Rev. B {\bf {59}}, 1758 (1999).

\bibitem{vasp}
G. Kresse and J. Furthm\"uller, Phys. Rev. B {\bf 54}, 11169 (1996).

\bibitem{key-1}J. A. Rodriguez, A. Etxeberria, L. Gonzalez, and A. Maiti,
J. Chem. Phys. {\bf 117}, 2699 (2002).

\bibitem{key-2}X. Ming, J.-W. Yin, X.-L. Wang, C.-Z.
Wang, Z.-F. Huang, and G. Chen, Solid State Sci. \textbf{12}, 938
(2010).

\bibitem{key-3}C. G. Van de Walle, D. B. Laks, G. F. Neumark, and
S. T. Pantelides, Phys. Rev. B \textbf{47}, 9425 (1993).


\bibitem{gap-err}
See e.g. R. W. Godby and P. Garcia-Gonzalez, in
{\it A Primer in Density Functional Theory}, Springer Lecture Notes in Physics,
 C. Fiolhais, F. Nogueira, M. Marques eds. (Springer, Berlin 2003), Ch. 5, p.185.

\bibitem{indiani}
H. K. Chandra, K. Gupta, A. K. Nandy, and P. Mahadevan, Phys. Rev. B {\bf 87}, 214110 (2013)

\end{thebibliography}
\end{document}